\documentclass[aps,pre,floats,twocolumn,showpacs,superscriptaddress]{revtex4}
\usepackage[dvips]{graphicx}
\newcommand{\comment}[1]{{}}

\newcommand{\beq}{\begin{equation}}
\newcommand{\eeq}{\end{equation}}
\newcommand{\ba}[1]{\begin{array}{#1}}
\newcommand{\ea}{\end{array}}
\newcommand{\bea}{\begin{eqnarray}}
\newcommand{\eea}{\end{eqnarray}}
\newcommand{\nn}{\nonumber \\}
\newcommand{\ben}{\begin{enumerate}}
\newcommand{\een}{\end{enumerate}}
\newcommand{\bit}{\begin{itemize}}
\newcommand{\eit}{\end{itemize}}
\newcommand{\bde}{\begin{description}}
\newcommand{\ede}{\end{description}}

\newcommand{\ds}{\displaystyle}

\begin{document}

\title{Analysis of community structure in networks of correlated data}

\author{Sergio G\'omez}
\affiliation{Departament d'Enginyeria Inform{\`a}tica i Matem{\`a}tiques,
  Universitat Rovira i Virgili,
  43007 Tarragona, Spain}

\author{Pablo Jensen}
\affiliation{Laboratoire de Physique, ENS Lyon, CNRS and Universit\'e de Lyon, IXXI - Institut des Syst\`emes Complexes,  5 rue du Vercors, 69007 Lyon, France}

\author{Alex Arenas}
\affiliation{Departament d'Enginyeria Inform{\`a}tica i Matem{\`a}tiques,
  Universitat Rovira i Virgili,
  43007 Tarragona, Spain}

\date{\today}

\begin{abstract}
We present a reformulation of modularity that allows the analysis of the community structure in networks of correlated data. The new modularity preserves the probabilistic semantics of the original definition even when the network is directed, weighted, signed, and has self-loops. This is the most general condition one can find in the study of any network, in particular those defined from correlated data. We apply our results to a real network of correlated data between stores in the city of Lyon (France).
\end{abstract}

\pacs{89.75.Hc, 02.10.Ox, 02.50.-r}

\maketitle

\section{Introduction}
Complex networks are graphs representative of the intricate connections between elements in many natural and artificial systems \cite{strogatz,havlin,barabasi,physrep}, whose description in terms of statistical properties have been largely developed looking for a universal classification of them. However, when the networks are locally analyzed, some characteristics that become partially hidden in the global statistical description emerge. The most relevant is perhaps the discovery in many of them of {\em community structure}, meaning the existence of densely (or strongly) connected groups of nodes, with sparse (or weak) connections between these groups \cite{firstnewman}.

The study of the community structure helps to elucidate the organization of the networks and, eventually, could be related to the functionality of groups of nodes \cite{amaral}. The most successful solutions to the community detection problem, in terms of accuracy and computational cost required, are those based in the optimization of a quality function called {\em modularity} proposed by Newman and Girvan \cite{newgirvan} that allows the comparison of different partitioning of the network. The extension of modularity to weighted \cite{wnewman} and directed networks \cite{mesoscales,leicht} has been the first steps towards the analysis of the community structure in general networks.

Very often networks are defined from correlation data between elements. The common analysis of correlation matrices uses classical or advanced statistical techniques \cite{song}. Nevertheless an alternative analysis in terms of networks is possible. The network approach usually consists in filtering the correlation data matrix, by eliminating poorly correlated pairs according to a threshold, and by keeping unsigned the value of the correlation, producing a network of positive links and no self-loops (self-correlations). Recently, some authors pointed out the possibility to analyze these networks via spectral decomposition \cite{heimo1,heimo2}. We devise also the possibility to analyze them in terms of Newman's modularity to reveal the community structure (clusters) of the correlated data. However, any of these approaches can be misleading because of two facts: first, the sign of the correlation is important to avoid the mixing of correlated and anti-correlated data, and second, the existence of self-loops is critical for the determination of the community structure \cite{mesoscales}. Here we propose a method to extract the community structure in networks of correlated data, that accounts for the existence of signed correlations and self-correlations, preserving the original information. To this end, we extend the modularity to the most general case of directed, weighted and signed links. We will show the performance of our method in a real network of correlations between commercial activities, previously analyzed in \cite{jensen} using a Potts model.

\section{Generalization of modularity}
Given an undirected network partitioned into communities, the modularity of a given partition is, up to a multiplicative constant, the probability of having  edges falling within groups in the network minus the expected probability in an equivalent (null case) network with the same number of nodes, and edges placed at random preserving the nodes' strength, where the strength of a node stands for the sum of the weights of its connections \cite{newanaly}. In mathematical form, modularity is expressed in terms of the weighted adjacency matrix $w_{ij}$, that represents the value of the weight in the link between $i$ and $j$ ($0$ if no link exists), as \cite{newanaly}
\beq
  Q = \frac{1}{2w} \sum_i\sum_j \left(
        w_{ij} - \frac{w_i w_j}{2w}
      \right) \delta(C_i,C_j) \,,
  \label{QW}
\eeq
where $C_i$ is the community to which node $i$ is assigned, the Kronecker delta function $\delta(C_i,C_j)$ takes the values, 1 if nodes $i$ and $j$ are into the same community, 0 otherwise, the strengths are $w_i=\sum_j w_{ij}$, and the total strength is $2w=\sum_i w_i =\sum_i \sum_j w_{ij}$.

The larger the modularity, the larger the deviation from the null case and the better the partitioning. Note that the optimization of the modularity cannot be performed by exhaustive search since the number of different partitions are equal to the Bell \cite{bell} or exponential numbers, which grow at least exponentially in the number of nodes $N$. Indeed, optimization of modularity is a NP-hard (Non-deterministic Polynomial-time hard) problem \cite{brandes}. Several authors have attacked the problem proposing different optimization heuristics \cite{newfast, clauset, rogernat, duch, pujol, newspect}.

\begin{figure}[!t]%figure1
  \begin{center}
  \begin{tabular}[b]{ccc}
    \raisebox{-25pt}{\includegraphics[width=130pt]{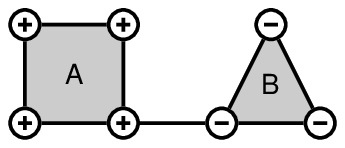}}
    &
    $\Rightarrow$\hspace{5pt}
    &
    \raisebox{-30pt}{\includegraphics[width=75pt]{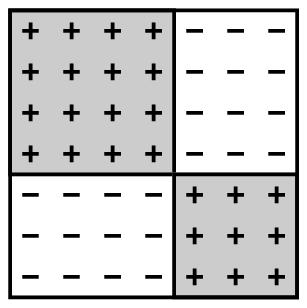}}
  \end{tabular}
  \end{center}
\caption{Network with well-defined community structure and its correlation matrix.}
\label{toy}
\end{figure}

To demonstrate the flaws of modularity when trying to extract the community structure of correlated data we show the following example. Suppose we have a network with a well defined community structure as the one presented in Fig.~\ref{toy}. Let us pretend that each community is indeed a functional community, in such a way that nodes in every group have different states. To simplify the mathematics we will consider that the nodes in community A are in a state $+1$, and nodes in community B are in a state $-1$. After, we define the correlation between these data as, for example, $R_{ij}=S_i S_j$, $S_i$ and $S_j$ being the corresponding states of nodes $i$ and $j$. The question is: can we infer communities A and B from the correlated data represented in matrix $R$? Applying modularity, the answer is negative. Let us sketch the proof. The matrix R is blockwise composed of submatrices $R_{AA}$, $R_{AB}$, $R_{BA}$, and $R_{BB}$. The blocks $R_{AA}$ and $R_{BB}$ are all valued $+1$, and $R_{AB}$ and $R_{BA}$ are valued $-1$. Any matrix of this form results in zero modularity for all partitions, since $R_{ij}=\frac{w_i w_j}{2w}$ for all pairs (see Eq.~\ref{QW}).

To reveal the community structure in the network presented in Fig.~\ref{toy} from
its correlation matrix, it is necessary to revise the formulation of modularity.
Let us suppose that we have a weighted undirected complex network with weights $w_{ij}$
as above. The relative strength $p_i$ of a node
\beq
  p_i = \frac{w_i}{2 w}\,,
\eeq
may be interpreted as the probability that this node makes links to other
ones, if the network were random. This is precisely the approach taken
by Newman and Girvan to define the modularity null case term, which reads
\beq
  p_i p_j = \frac{w_i w_j}{(2 w)^2}\,.
\eeq

The introduction of negative weights destroys this probabilistic interpretation
of $p_i$, since in this case the values of $p_i$ are not guaranteed to be between
zero and one. The problem is the implicit hypothesis that there is only one
unique probability to link nodes, which involves both positive and negative
weights. To solve this problem, we have to introduce two different probabilities to
form links, one for positive and the other for negative weights.

Let us formalize this approach. First, we separate the positive and negative
weights:
\beq
  w_{ij} = w_{ij}^{+} - w_{ij}^{-}\,,
\eeq
where
\bea
  w_{ij}^{+} & = & \max\{0,  w_{ij}\}\,, \\
  w_{ij}^{-} & = & \max\{0, -w_{ij}\}\,.
\eea
The positive and negative strengths are given by
\bea
  w_i^{+} & = & \sum_j w_{ij}^{+}\,, \\
  w_i^{-} & = & \sum_j w_{ij}^{-}\,,
\eea
and the positive and negative total strengths by
\bea
  2 w^{+} & = & \sum_i w_i^{+} = \sum_i \sum_j w_{ij}^{+}\,, \\
  2 w^{-} & = & \sum_i w_i^{-} = \sum_i \sum_j w_{ij}^{-}\,.
\eea
Obviously,
\beq
  w_i = w_i^{+} - w_i^{-}
\eeq
and
\beq
  2 w = 2 w^{+} - 2 w^{-}\,.
\eeq

With these definitions at hand, the connection probabilities with
positive and negative weights are respectively
\bea
  p_i^{+} & = & \frac{w_i^{+}}{2 w^{+}}\,, \\
  p_i^{-} & = & \frac{w_i^{-}}{2 w^{-}}\,.
\eea
%The error would be to merge them into a single pseudo-probability
%\beq
%  \frac{w_i}{2 w} = \frac{w_i^{+} - w_i^{-}}{2 w^{+}- 2 w^{-}}\,.
%\eeq

Now, there are two terms which contribute to modularity: the first
one takes into account the deviation of actual positive weights
against a null case random network given by probabilities $p_i^{+}$,
and the other is its counterpart for negative weights. Thus, it
is useful to define
\bea
  Q^{+} & = & \frac{1}{2w^{+}} \sum_i \sum_j \left(
                w_{ij}^{+} - \frac{w_i^{+} w_j^{+}}{2w^{+}}
              \right) \delta(C_i,C_j)\,,
  \\
  Q^{-} & = & \frac{1}{2w^{-}} \sum_i \sum_j \left(
                w_{ij}^{-} - \frac{w_i^{-} w_j^{-}}{2w^{-}}
              \right) \delta(C_i,C_j)\,.
\eea

The total modularity must be a trade off between the tendency of
positive weights to form communities and that of negative weights
to destroy them. If we want that $Q^{+}$ and $Q^{-}$ contribute
to modularity proportionally to their respective positive and negative
strengths, the final expression for modularity $Q$ is
\beq
  Q = \frac{2 w^{+}}{2 w^{+} + 2 w^{-}} Q^{+} -
      \frac{2 w^{-}}{2 w^{+} + 2 w^{-}} Q^{-}\,.
\eeq
An alternative equivalent form for modularity $Q$ is
\bea
  Q = \frac{1}{2w^{+} + 2 w^{-}} & \ds \sum_i \sum_j &
        \left[ w_{ij} - \left(
          \frac{w_i^{+} w_j^{+}}{2w^{+}}- \frac{w_i^{-} w_j^{-}}{2w^{-}}
        \right) \right] \nn
        &&\times \delta(C_i,C_j)\,.
  \label{QWS}
\eea

The main properties of Eq.~(\ref{QWS}) are the following: without negative weights, the
standard modularity is recovered; modularity is zero when all nodes are together
in one community; and it is antisymmetric in the weights, i.e.
$Q(C,\{w_{ij}\}) = - Q(C,\{-w_{ij}\})$\,.

The extension to directed networks \cite{njp} is simply obtained by the substitutions
\bea
  w_{i}^{\pm} &\rightarrow &  w_{i}^{\pm,\mbox{\scriptsize out}} = \sum_{k} w_{ik}\,,\\
  w_{j}^{\pm} &\rightarrow &  w_{j}^{\pm,\mbox{\scriptsize in}} = \sum_{k} w_{kj} \,.
\eea

\section{Comparison with other methods}
\begin{figure}[!t]%figure2
  \begin{center}
    \includegraphics[width=200pt]{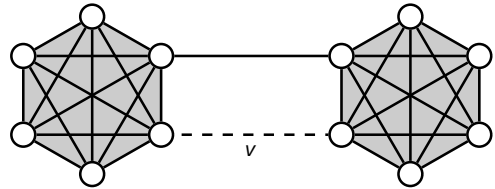}
  \end{center}
\caption{Network with two well-defined communities. Solid lines correspond to positive links, and the dashed line to the only negative link, with weight $v<0$.}
\label{hexa}
\end{figure}

In Fig.~\ref{hexa} we show a simple example of a network for which the original Newman modularity Eq.~(\ref{QW}) and the Potts model in \cite{jensen} do not yield the expected partition in two communities, whereas our new modularity Eq.~(\ref{QWS}) succeeds. It consists in two cliques, formed by positive links, and connected by two edges, one positive and the other negative. All positive links have a weight $+1$, and the negative a weight $v<0$. Any size of the cliques greater than or equal to three does the job.

First, the Potts model in \cite{jensen} is based on a Hamiltonian which only takes into account the difference between positive and negative weights within the modules, and is equivalent to modularity but without the null case term. In the network Fig.~\ref{hexa}, if $|v|<1$, the strength between the two cliques is $1+v>0$, thus the Potts model is rewarded to join both cliques in the same module. Clearly, the absence of the null case is responsible of this incorrect result.

On the other hand, the original definition of modularity (Eq.~\ref{QW}), which does include a null case, was not designed to cope with negative weights. In this example, its optimal partition is again a single module containing all the nodes if the value of $|v|$ is greater than the number of positive links.

In \cite{traag} the authors propose an alternative definition of modularity for positive and negative links. Their work, also based on a Potts model representation of the network communities' assignment \cite{bornholdt}, is totally compatible with the definition found in the current work, and equivalent for the values of their parameters $\lambda = \gamma = 1$.

\section{Application to a real network}
We now turn to an example of community structure detection using our method in a specific social network. We deal with the spatial distribution of retail activities in the city of Lyon, thanks to data obtained at the Lyon's Commerce Chamber \footnote{The Commerce Chamber classifies retail activities according to commercial criteria derived from an experienced knowledge of the field. This classification is adopted here as reference.}. We have shown in \cite{jensen} how to transform data on locations into a matrix of correlated data, in this case of attractions/repulsions (i.e.\ positive and negative links) between retail activities. To compute the interaction between activities A and B, the idea is to compare the concentrations of B stores in the neighborhood of A stores to a reference concentration obtained by locating the B stores randomly. To compute the random reference, the idea \cite{duranton} is to locate the B stores on the array of {\em all existing} store sites. This is the best way to take into account automatically the geographical peculiarities of each town. The logarithm of the ratio of the actual concentration to the reference concentration gives the interaction coefficient, which is positive for attractions and negative for repulsions, as anticipated.

More precisely, the (self) interaction of $N_{A}$ A stores embedded in a larger set of $N_t$ locations is
\beq
  \label{com-aaa}
  a_{AA}(r) =  \log_{10} \frac{N_t - 1}{N_A (N_A-1)} \sum_{i=1}^{N_A} \frac{ N_A
  (A_i)}{N_t (A_i)} \,,
\eeq
where $N_A (A_i)$ and ${N_t (A_i)}$ represent the number of A stores and the
total number of stores in the neighborhood of store $A_i$, i.e. locations at a
distance smaller than {$r$}. Similarly, the coefficient characterizing the
spatial distribution of the $B_i$ around the $A_i$ is
\beq
  \label{com-aab}
  a_{AB}(r) = \log_{10} \frac{N_t - N_A}{N_A N_B} \sum_{i=1}^{N_A} \frac{ N_B
  (A_i)}{N_t
  (A_i) - N_A (A_i)} \,,
\eeq
where $N_A (A_i)$, $N_B (A_i)$ and $N_t (A_i)$ are respectively the $A$, $B$ and total number of locations in the neighborhood of point $A_i$ (not counting $A_i$). Both $a_{AA}$ and $a_{AB}$ are defined so that they take value 0 when there are no spatial correlations. In the case of the $a_{AB}$ coefficient, this means that the local $B$ spatial concentration is not perturbed, on average, by the presence of A stores, and is equal to the average concentration over the whole town, $\frac{N_B}{N_t-N_A}$. Only coefficients which deviate significantly from 0, using a Montecarlo sampling, are taken into account in the adjacency matrix. The final result of the analysis of the 11629 stores in Lyon is a directed network with 97 nodes (retail activities) and 1131 links, 715 positive and 416 negative.

We analyze the community structure of the resulting network using the modularity defined in Eq.~(\ref{QWS}). The optimization method used is Tabu search \cite{mesoscales} that for this case gave the highest modularity when compared to others \cite{jstat}. We perform a comparison between the different partitions obtained optimizing independently Eq.~(\ref{QW}) (resulting in 4~communities) and Eq.~(\ref{QWS}) (resulting in 6~communities), against the Lyon's Commerce Chamber retail activities classification (9~communities predefined). The similarity of the first two partitions to the third one is measured using three different indices, namely the Rand Index \cite{rand}, the Jaccard Index \cite{jaccard}, and the Normalized Mutual Information (NMI) \cite{strehl} (see Table~\ref{tab}). The larger their values, the more similar the partitions are. All indices show a better performance of Eq.~(\ref{QWS}) in recovering the actual communities provided by the  Lyon's Commerce Chamber. Note that in both modularities we have used all the positive and negative links. Therefore, the increase in performance can only be attributed to a proper use of the information embedded in the links.

\begin{table}[t]
\caption{Comparison between the different partitions and the Lyon Chamber of Commerce classification.}
\begin{tabular}{lcc}
\hline
\hline
& optimal partition & optimal partition\\
& of Eq.~(\ref{QW}) & of Eq.~(\ref{QWS}) \\
\hline
Rand Index    & 0.6168 & 0.6952\\
Jaccard Index & 0.1336 & 0.1426\\
NMI           & 0.1458 & 0.2310\\
\hline
\end{tabular}
\label{tab}
\end{table}

Our method is also helpful to understand the spatial organisation of retail stores. To interpret the information conveyed by the network links, we use of the z-score \cite{rogernat}. The basic idea consists in computing the z-score (Z) of the internal strength of each node with respect to the average internal strength of the community to which is assigned. To be consistent with our approach along the paper both quantities should be evaluated consistently with the sign of the interactions and with the directionality of links, then
\begin{equation}
  Z^{\pm,\mbox{\scriptsize in/out}}_{i} =
    \frac{w^{\pm,\mbox{\scriptsize in/out}}_{i,\mbox{\scriptsize int}} -
           \langle w^{\pm,\mbox{\scriptsize in/out}}_{\mbox{\scriptsize int}}\rangle}
         {\sigma(w^{\pm,\mbox{\scriptsize in/out}}_{\mbox{\scriptsize int}})} \,,
\label{z}
\end{equation}

\noindent where subindices `int' express that links are restricted within the community to which node $i$ belongs to, `in/out' refer to the direction of links, and $\langle\cdots\rangle$ and $\sigma$ are the average and standard deviation of the corresponding variables, respectively.

Using the z-score we can answer some questions about the role of nodes in their communities. For example, one can study, for each community, which are the most attractive retailers (max $Z^{+,\mbox{\scriptsize out}}$), the most repulsive retailers (max $Z^{-,\mbox{\scriptsize out}}$), the most attracted retailers (max $Z^{+,\mbox{\scriptsize in}}$), and the most repelled retailers (max $Z^{-,\mbox{\scriptsize in}}$). In Table~\ref{tab2} we show the three highest results of these z-scores obtained for the largest community found (34 retail activities). This group gathers the proximity stores, which means mainly food stores. Here are some examples of the understanding of the spatial organisation of retail stores allowed by our method. Sports facilities and funeral services are peculiar because they strongly attract (and are attracted) by some specific activities that go along with them almost systematically, e.g.\ car repairs and small hardware stores. Gas stations enjoy a paradoxical situation in this group, since they represent the most attracted and the most repelled activity. There is an interesting commercial interpretation of this paradox: gas stations tend to have the most specific commercial environment, strongly attracting some of the group's activities (such as supermarkets) and being strongly repelled by others which however are in the proximity store group (for example, butchers or cake shops stores almost never have gas stations close to them). Dairy products and cake shops strongly repel some specific of the activities that belong to their same group, such as car repairs or firm's restaurants.

% It is very significant the situation found for gas stations, the data tell us that gas retailers tend to have their location close to the rest of retailers in the community, while retailers do not want to have a gas station close to them. The case of sport facilities is also interesting to mention, they tend to have their location close to the rest of retailers and at the same time are very welcomed to be close. Dairy products shops and cake shops, tend to isolate from the rest of retailers, and Flea markets are repelled by the retailers within the community. Curiously, funeral services are centrally situated in the city and are welcomed by the retailers of its community.

\begin{table}[t]
\caption{Roles of retailers within communities.}
\begin{tabular} {ccccc}
\hline
\hline
&+ attractive & + repulsive& + attracted & + repelled \\
%& out+ & out- & in+ & in- \\
\hline
%& Gas Station & Dairy products & Funeral Services & Gas Station \\
%& Sports facility & Cake shop & Sports facility & Flea market \\
& Funeral Services & Dairy products & Gas Station & Gas Station \\
& Sports facility & Cake shop & Sports facility & Flea market \\
& Car dealer & Drugstore & Funeral Services & Car dealer \\
\hline
\end{tabular}
\label{tab2}
\end{table}

\section{Conclusions}
Summarizing, we have proposed a new formulation of modularity that allows for the analysis of any complex network, in general with links directed, weighted, signed and with self-loops, preserving the original probabilistic semantics of modularity. With this definition one can analyze networks arising from correlated data without necessarily symmetrizing the network, skipping auto-correlation or considering only the unsigned value of the correlations. We devise that other methods are also likely to be appropriate for this task, after its pertinent adaptation, for example the analysis via clique percolation \cite{palla}, or specifically methods based on the minimization of the energy function of an equivalent spin glass system, were weighted signed links can be interpreted in terms of ferromagnetic and anti-ferromagnetic interactions between spins \cite{bornholdt}.

We have analyzed within the scope of the new modularity an interesting model of attraction-repulsion of retail stores in a large city, previously reported in \cite{jensen}. The results overcome those obtained using the original definition of modularity when compared to the Lyon Chamber of Commerce classification, and also point out the necessity of defining new roles of nodes based on directionality and sign of the weights of links, as we have proposed for the z-score.

\begin{acknowledgments}
  We thank J. Borge for help with the simulations. This work has been partially supported by the Spanish DGICYT Project FIS2006-13321-C02-1. We gratefully acknowledge Christophe Baume and Bernard Gagnaire from Lyon's Commerce Chamber who have kindly provided the location data.
\end{acknowledgments}


\begin{thebibliography}{99}

\bibitem{strogatz}
S. H. Strogatz, {\em Nature} {\bf 410}, 268 (2001).

\bibitem{havlin}
C. M. Song, S. Havlin, H. A. Makse, {\em Nature} {\bf 433}, 392 (2005).

\bibitem{barabasi}
A.-L. Barab{\'a}si, {\em Science} {\bf 308}, 639 (2005)

\bibitem{physrep}
S. Boccaletti, V. Latora, Y. Moreno, M. Chavez, D.-U. Hwang, {\em Phys. Rep.} {\bf 424}, 175 (2006).

\bibitem{firstnewman}
M. Girvan, M. E. J. Newman, {\em Proc. Natl. Acad. Sci. USA} {\bf 99}, 7821 (2002).

\bibitem{amaral}
R. Guimer{\`a}, L. A. N. Amaral, {\em Nature} {\bf 433}, 895 (2005).

\bibitem{newgirvan}
M. E. J. Newman, M. Girvan, {\em Phys. Rev. E} {\bf 69}, 026113 (2004).

\bibitem{wnewman}
M. E. J. Newman, {\em Phys. Rev. E} {\bf 70}, 056131 (2004).

\bibitem{mesoscales}
A. Arenas, A. Fern\'andez, S. G\'omez, {\em New J. Phys.} {\bf 10}, 053039 (2008).

\bibitem{leicht}
E. A. Leicht, M. E. J. Newman, {\em Phys. Rev. Lett.} {\bf 100}, 118703 (2008).

\bibitem{song}
P. X.-K Song, Correlated Data Analysis: Modeling, Analytics, and Applications, (Springer Series in Statistics, New York, 2007).

\bibitem{heimo1}
T. Heimo, J. Saram\"aki, J.-P. Onnela, K. Kaski, {\em Physica A} {\bf 383}, 147 (2007).

\bibitem{heimo2}
T. Heimo, G. Tib\'ely, J. Saram\"aki, K. Kaski, J. Kert\'esz, {\em Physica A} {\bf 387}, 5930-5945 (2008).

\bibitem{jensen}
P. Jensen, {\em Phys. Rev. E} {\bf 74}, 035101 (2006)

\bibitem{newanaly}
M. E. J. Newman, {\em Phys. Rev. E} {\bf 70}, 056131 (2004).

\bibitem{bell}
E. T. Bell, {\em Amer. Math. Monthly} {\bf 41}, 411 (1934).

\bibitem{brandes}
U. Brandes, D. Delling, M. Gaertler, R. Goerke, M. Hoefer, Z. Nikoloski, D. Wagner, {\em IEEE Transactions on Knowledge and Data Engineering} {\bf 20(2)}, 172 (2008).

\bibitem{newfast}
M. E. J. Newman, {\em Phys. Rev. E} {\bf 69}, 066133 (2004).

\bibitem{clauset}
A. Clauset, M. E. J. Newman, C. Moore, {\em Phys. Rev. E} {\bf 70}, 066111 (2004).

\bibitem{rogernat}
R. Guimer{\`a}, L. A. N. Amaral, {\em J. Stat. Mech.}, P02001 (2005).

\bibitem{duch}
J. Duch, A. Arenas, {\em Phys. Rev. E} {\bf 72}, 027104 (2005).

\bibitem{pujol}
J. M. Pujol, J. B{\'e}jar, J. Delgado, {\em Phys. Rev. E} {\bf 74}, 016107 (2006).

\bibitem{newspect}
M. E. J. Newman, {\em Proc. Natl. Acad. Sci. USA} {\bf 103}, 8577 (2006).

\bibitem{njp}
A. Arenas, J. Duch, A. Fern\'andez, S. G\'omez, {\em New J. Phys.} {\bf 9}, 176 (2007).

\bibitem{traag}
V. A. Traag, J. Bruggeman, arXiv:0811.2329 (2008).

\bibitem{bornholdt}
J. Reichardt, S. Bornholdt, {\em Phys. Rev. E} {\bf 74}, 016110 (2006).

\bibitem{duranton}
G. Duranton, H. G. Overman, {\em The Review of Economic Studies} {\bf 72}, 1077 (2005).

\bibitem{jstat}
L. Danon, A. D\'iaz-Guilera, J. Duch, A. Arenas, {\em J. Stat. Mech.}, P09008 (2005).

%\bibitem{jensenmichel}
%P. Jensen, J. Michel, {\em Annals of Regional Science}, submitted.

\bibitem{rand}
W. M. Rand, {\em J. Am. Stat. Assoc.} {\bf 66}, 846 (1971).

\bibitem{jaccard}
P. Jaccard, {\em The New Phytologist} {\bf 11(2)}, 37 (1912).

\bibitem{strehl}
A. Strehl, J. Ghosh, {\em J. Machine Learning Research} {\bf 3}, 583 (2002).

\bibitem{palla}
G. Palla, I. Der{\'e}nyi, I. Farkas, T. Vicsek, {\em Nature} {\bf 435}, 814 (2005).

\end{thebibliography}
\end{document}